\documentclass[aps,prl,onecolumn,superscriptaddress]{revtex4}

\usepackage{graphicx}
\usepackage{dcolumn}
\usepackage{bm}
\usepackage {amsmath}

\begin{document}

\title{Highly efficient entanglement swapping and teleportation at telecom wavelength}

\author{Rui-Bo Jin}
\email{ruibo@nict.go.jp}
\affiliation{National Institute of Information and Communications Technology (NICT), 4-2-1 Nukui-Kitamachi, Koganei, Tokyo 184-8795, Japan}

\author{Masahiro Takeoka}
\affiliation{National Institute of Information and Communications Technology (NICT), 4-2-1 Nukui-Kitamachi, Koganei, Tokyo 184-8795, Japan}

\author{Utako Takagi}
\affiliation{National Institute of Information and Communications Technology (NICT), 4-2-1 Nukui-Kitamachi, Koganei, Tokyo 184-8795, Japan}
\affiliation{Tokyo University of Science, 1-3 Kagurazaka, Shinjuku-ku, Tokyo 162-8601, Japan }

\author{Ryosuke Shimizu}
\affiliation{University of Electro-Communications, 1-5-1 Chofugaoka, Chofu, Tokyo 182-8585, Japan}

\author{Masahide Sasaki}
\affiliation{National Institute of Information and Communications Technology (NICT), 4-2-1 Nukui-Kitamachi, Koganei, Tokyo 184-8795, Japan}

\date{\today }

\begin{abstract}

Entanglement swapping at telecom wavelengths is at the heart of quantum networking in optical fiber infrastructures.
Although entanglement swapping has been demonstrated experimentally so far using various types of entangled photon sources
both in near-infrared and telecom wavelength regions, the rate of swapping operation has been too low to be applied
to practical quantum protocols, due to limited efficiency of entangled photon sources and photon detectors.
Here we demonstrate drastic improvement of the efficiency at telecom wavelength by using two ultra-bright entangled photon sources and four highly efficient superconducting nanowire single photon detectors.
We have attained a four-fold coincidence count rate of 108 counts per second, which is three orders higher than the previous experiments
at telecom wavelengths. A raw (net) visibility in a Hong-Ou-Mandel interference between the two independent entangled sources was 73.3 $\pm$ 1.0\% (85.1 $\pm$ 0.8\%).
We performed the teleportation and entanglement swapping, and obtained a fidelity of 76.3\% in the swapping test.
Our results on the coincidence count rates are comparable with the ones ever recorded in teleportation/swaping and multi-photon entanglement generation experiments at around 800\,nm wavelengths.
Our setup opens the way to practical implementation of device-independent quantum key distribution and its distance extension by the entanglement swapping
as well as multi-photon entangled state generation in telecom band infrastructures with both space and fiber links.

\end{abstract}

\maketitle

\section{Introduction}

Distribution of quantum entanglement through optical channels is the basis of implementing quantum information and communication
protocols, which do not have any classical counterparts, such as device-independent quantum key distribution (DI-QKD) \cite{Acin2007, Ekert1991},
quantum secret sharing \cite{Tittel2001, Gisin2002}, quantum repeater network \cite{Briegel1998, Dur1999, Duan2001, Sangouard2008,  Kaltenbaek2009}, and so on.
The distance of direct transmission of entanglement is, however, severely limited, because the entanglement is easily destroyed by the channel loss and channel noises.
Extending the distance and networking the entanglement requires entanglement swapping as the very elementary protocol.
This is to convert two independent entangled photon pairs, say, photons A and B, and C and D, to a new entangled pair of photons between A and D,
those are not originally entangled, by performing a  Bell measurement on photons B and C.

Practical methods for the swapping at present is to prepare entangled photons from spontaneous parametric down conversion (SPDC),
to detect the two photons at the intermediate node by two single-photon detectors, and to herald the success event for the swapping.
Thus the protocol is probabilistic. Its success rate is directly determined by the four-fold coincidence count (4-fold CC) rate.

The first entanglement swapping experiment was carried out in 1998, with SPDC process in BBO crystal \cite{Pan1998}.
Since then, many proof-of-principle experiments have been demonstrated at near infra-red wavelength range \cite{Pan1998, Jennewein2001, Kaltenbaek2009}.
Demonstrations at telecom wavelengths have also been demonstrated with SPDC in bulk crystals \cite{Marcikic2003, Riedmatten2005, Wu2013},
in waveguides \cite{ Halder2007, Xue2010,  Xue2012}, or with spontaneous four-wave mixing (SFWM) in fibers \cite{Takesue2009}.
Unfortunately, however, the efficiencies were very low, especially at telecom wavelengths so far.
For example, the maximum 4-fold CC rate has been 0.08 counts per second (cps) \cite{Wu2013}.
This limits practical applications of entanglement swapping to quantum information and communication protocols.

In this work, we demonstrate highly efficient entanglement swapping by utilizing our high-quality entangled photon source \cite{Jin2014OE}
and highly efficient superconducting nanowire single photon detectors (SNSPDs) \cite{Miki2013, Yamashita2013}.
In our experiment, 4-fold CC rate of 108 cps was attained, which is three orders higher than the previous record \cite{Wu2013}.
A net visibility is 85.1 $\pm$ 0.8\% in Hong-Ou-Mandel interference between two independent entangled sources.
We also demonstrate high quality teleportation experiment, with 2-fold CC of 150 kcps, which is comparable to those obtained in highly efficient schemes
in the near-infrared wavelengths \cite{Herbst2014, Yin2012, Yao2012, Huang2011}, and with the entanglement visibility of 98\,\%, which is the highest among \cite{Herbst2014, Yin2012, Yao2012, Huang2011}.

\section{Experiment and results }

\begin{figure}[tbp]
\includegraphics[width= 0.85 \textwidth]{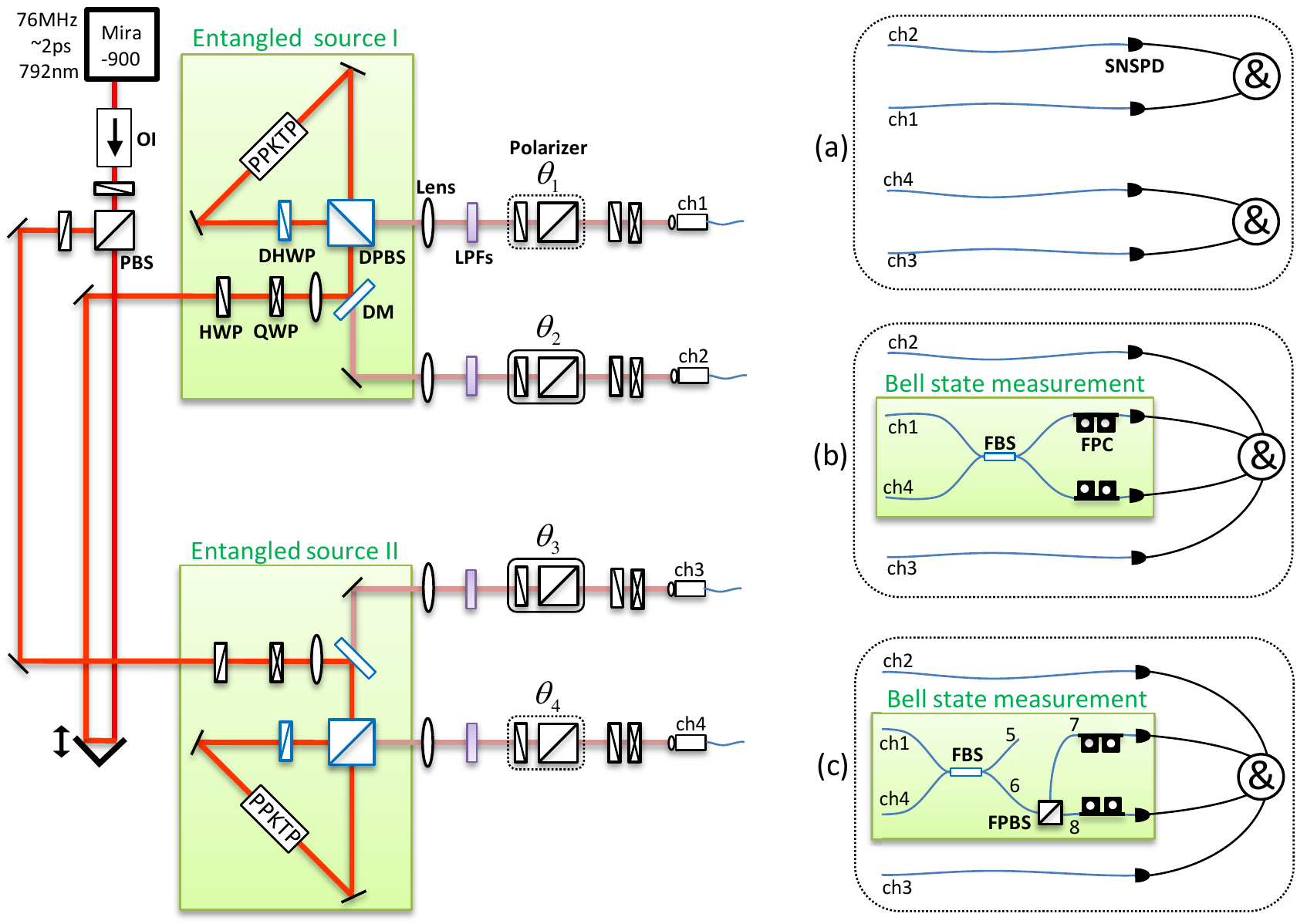}
\caption{  The experimental setup. Picosecond laser pulses (76 MHz, 792 nm, temporal duration $\sim$ 2 ps) from a mode-locked Titanium sapphire laser (Mira900, Coherent Inc.) were divided into two paths, and pumped two entangled photon sources which were in Sagnac-loop configurations and  focused by two $f = 200$ mm lenses (beam waist $\sim$ 45 $\mu$m). Each Sagnac-loop consisted of a dual-wavelength polarization beam splitter (DPBS), a dual-wavelength half-wave plate (DHWP), and a 30-mm-long PPKTP crystal with a polling period of 46.1 $\mu$m for a type-II collinear group-velocity-matched SPDC. The downcoverted photons passed through four sets of longpass filters (LPF), four polarizers ($\theta_1$ to $\theta_4$), four sets of half-wave plate (HWP) and quarter-wave plate (QWP), and then coupled into four channels (ch1 to ch4) of single-mode fibers.   Two sets of fiber polarization controllers (FPC) were used to optimize the polarization of the photons  before detected by SNSPDs, which were connected to a coincidence counter (\&).
$\theta_1$ and $\theta_4$ were  removed in the tests of teleportation and entanglement swapping, so as to couple both the horizontal (H) and vertical (V) polarized photons into the fiber beam splitter (FBS, Thorlabs, PBC1550SM-FC ). (a) is for entangled source test. (b) is for Hong-Ou-Mandel interference and teleportation  tests. (c) is for entanglement swapping test. In the teleportation test in (b), Bell state measurement (BSM) was realized by a FBS.  To overcome the polarization dependance of the SNSPDs in entanglement swapping test in (c), the BSM was realized by using the combination of a FBS and a fiber based PBS (FPBS, PBC1550SM-FC, Thorlabs), which was constituted of a calcite prism and  input/output fibers.
 Four coarse bandpass filters (not shown) were inserted in  four channels (ch1 - ch4) for the tests of teleportation and entanglement swapping.
 } \label{setup}
\end{figure}

The experimental setup is shown in Fig. \ref{setup}.
The entangled photon source is based on a SPDC from a group-velocity-matched periodically poled KTiOPO$_4$ (GVM-PPKTP) crystal in a Sagna-loop configuration  \cite{Jin2014OE}.
This entangled photon source has a spectral purity as high as 0.82 \cite{Jin2013OE},  which can widen applications with multi-source of entangled photons.
The SNSPD has a  maximum system detection efficiency (SDE) of 79 \% with a dark count rate (DCR) of 2 kcps.
For more details of this Saganc-GVM-PPKTP entangled source, refer to Refs. \cite{Jin2014OE, Kim2006, Takeoka2014}.
By carefully improving the coupling efficiency, we have improved the coincidence counts from 20 kcps (in Ref.\cite{Jin2014OE}) to 40 kcps, at 10 mW pump power.
The overall efficiency was estimated as 0.20.
Further, the minimal interference visibility in polarization correlation measurement  was improved from  $\sim$ 96\% (in Ref.\cite{Jin2014OE}) to  $\sim$ 98\% at 10 mW pump power, by finely optimizing the alignment in the Sagnac-loop.

\subsection{The entangled source }

\begin{figure*}[tbp]
\includegraphics[width= 0.75 \textwidth]{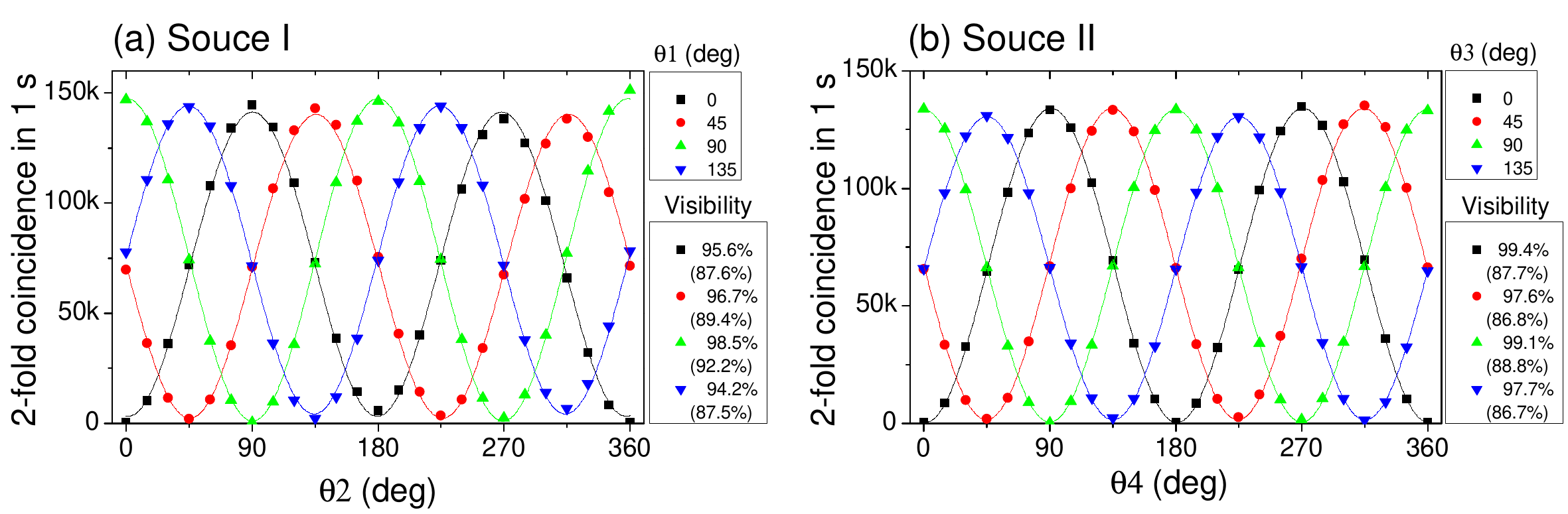}
\caption{ Two-fold coincidence counts in one second as a function of the two polarizers, with
a pump power of 80 mW for entangled source I  (a) and  85 mW for entangled source II (b). The background subtracted visibility (raw visibility) are
shown on the right side.   }
\label{EntangleSource}
\end{figure*}

First, we perform the photon polarization correlation measurement with the setup shown in  Fig. \ref{setup}(a).
With  pump powers of 80 mW for  source I and 85 mW for source II, we achieve coincidence counts of around 150 kcps, as shown in Fig. \ref{EntangleSource}, corresponding to 300 kcps without the polarizers. The corresponding mean photon numbers per pulse are around 0.1 for both sources. The raw visibilites are around 87\%, while the background subtracted visibilities (net visibilities) are around 98\% for each polarizers set for source II. The degradation of this visibility at high pump power is mainly caused by the multi-pair emission.
These high-brightness entangled photon sources guarantee the high count rate in the following teleportation and entanglement swapping experiments.

\subsection{Hong-Ou-Mandel interference}
\begin{figure*}[tbp]
\includegraphics[width= 0.7 \textwidth]{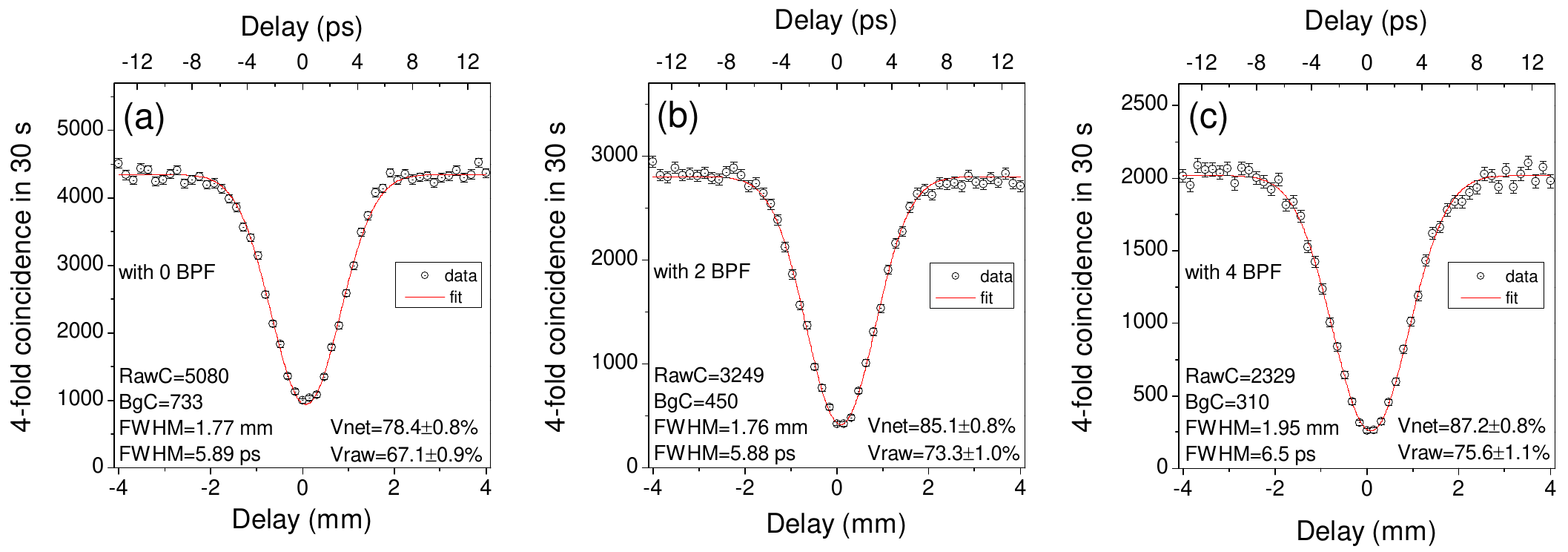}
\caption{ Hong-Ou-Mandel interference. (a) with no BPF, (b) with two BPFs, (c) with four BPFs. The error bars were added by assuming Poissonian statistics of these coincidence counts. }
\label{HOMI}
\end{figure*}

Next, we measure  a four-fold  Hong-Ou-Mandel (HOM) interference \cite{Hong1987} between ch1 and ch4 (heralded by ch2 and ch3 ) with 4 polarizers inserted in each channel in  Fig. \ref{setup}.
The polarization angles for $\theta_1$-$\theta_4$  are set at 0$^{\circ}$/90$^{\circ}$/90$^{\circ}$/0$^{\circ}$, respectively.
Firstly, we test the HOM interference with no bandpass filters inserted, whose result is  shown in  Fig. \ref{HOMI}(a).
The 4-fold CC is 169 cps (5080 counts in 30 s) and the raw visibility is 67.1 $\pm$ 0.9\%.
The background counts was mainly caused by the multi-pair emission in SPDC.
We subtract the background counts using the same method as shown in Refs \cite{Jin2013PRA}.
We block  only ch1 and measure 4-fold CCs, then block only ch4, and measure 4-fold CCs.
The sum of these two coincidence counts  constitutes the background count.
After background subtraction, the net visibility is 78.4 $\pm$ 0.8\%, which is consistent with our previous results in  Ref \cite{Jin2013PRA}.

To further increase the visibility, we should improve the spectral purity of photons.
As reported in Ref \cite{Jin2013OE}, the downconverted photons from our PPKTP crystal have an intrinsic spectral purity of 0.82, and this value can be improved to unity by inserting coarse bandpass filters (C-BPFs) to cut the side lobes in the joint spectral amplitude.
In this experiment, we prepared four C-BPFs which have near-Gaussian shape with FWHM (full width at half maximum) of 2.1 nm and peak transmission efficiency  of 93\% at the central wavelength of 1584 nm.
The overall transmittance efficiency of the C-BPFs is around 77 \%, tested with our downconverted photons , which
have spectra of Gaussian shape with  FWHM of 1.2 nm and center wavelength of 1584 nm.
We insert two C-BPFs in ch1 and ch4 and repeat the HOM interference, whose result is shown in  Fig. \ref{HOMI}(b).
The 4-fold CC rate is 108 cps (3249 counts in 30 s) and the raw  visibility is 73.3 $\pm$ 1.0\%.
After subtracting the  background multi-photon emission, the net visibility is 85.1 $\pm$ 0.8\%.
The raw (net) visibility was  improved by 6.2\% (6.7\% ) after the insertion of the two C-BPFs, due to the increase of the spectral purity.
However, this visibility improvement is lower than our expectation, which may be caused by the following reasons:
the photons generated from two different nonlinear crystals may have different spectral properties;
the transmission profiles of the BPFs can not be perfectly the same, which may also lead to the difference of the transmitted photons;
a small portion of  the side-lodes may be not filtered because the transmission shape of the BPFs is not in a perfect Gaussian-shape.

We also investigate the HOM interference with four C-BPFs inserted in each channel, whose result is shown in Fig. \ref{HOMI}(c).
The 4-fold CC rate is 78 cps (2329 in 30 s) and the raw (net)  visibility is 75.6 $\pm$ 1.1\% (87.2 $\pm$ 0.8\%).
The raw (net) visibility was  improved by 2.3\% (2.1\% ) after the insertion of the two more BPFs.
In the following test of quantum teleportation and entanglement swapping, all these four C-BPFs are inserted.

\subsection{Quantum teleportation}
After the test of Hong-Ou-Mandel interference, we remove the two PBSs in Polarizer 1 ($\theta _1$) and Polarizer 4 ($\theta _4$), then the setup is ready for the test of quantum teleportation.
The principle of teleportation in our experiment is as follow.
Assume photons in ch1 are in an initial state of $ \left| i \right\rangle _1  = (\alpha \left| H \right\rangle  + \beta \left| V \right\rangle )_1 $,  and photons in ch3 and ch4 are in an entangled state of $\left| {\psi ^ -  } \right\rangle _{34}  = \frac{1}{{\sqrt 2 }}(\left| {HV} \right\rangle  - \left| {VH} \right\rangle )_{34}$.
A partial Bell state measurement on photons in ch1 and ch4 will teleport the state $\left| i \right\rangle _1$  to  the photons in ch3, i.e., the final state of the photons in ch3 will be $ \left| f \right\rangle _3  = (\alpha \left| H \right\rangle  + \beta \left| V \right\rangle )_3 $.
This process can be expressed as \cite{Bouwmeester1997}:
\begin{equation}\label{eq0}
\begin{array}{l}
 \left| {\psi ^ -  } \right\rangle _{34}  \otimes \left| i \right\rangle _1  = \frac{1}{{\sqrt 2 }}(\left| {HV} \right\rangle  - \left| {VH} \right\rangle )_{34} (\alpha \left| H \right\rangle  + \beta \left| V \right\rangle )_1  \\
  = \frac{1}{2}[\left| {\psi ^ +  } \right\rangle _{41} ( - \alpha \left| H \right\rangle  + \beta \left| V \right\rangle )_3  + \left| {\psi ^ -  } \right\rangle _{41} (\alpha \left| H \right\rangle  + \beta \left| V \right\rangle )_3  + \left| {\phi ^ +  } \right\rangle _{41} (\alpha \left| V \right\rangle  - \beta \left| H \right\rangle )_3  + \left| {\phi ^ -  } \right\rangle _{41} (\alpha \left| V \right\rangle  + \beta \left| H \right\rangle )_3 ] \\
 \end{array}
\end{equation}
where $\left| {\psi ^ \pm  } \right\rangle  = \frac{1}{{\sqrt 2 }}(\left| {HV} \right\rangle  \pm \left| {VH} \right\rangle )$ and
$\left| {\phi ^ \pm  } \right\rangle  = \frac{1}{{\sqrt 2 }}(\left| {HH} \right\rangle  \pm \left| {VV} \right\rangle )$
are the four Bell states.
We only focus the second term:
\begin{equation}\label{eq1}
 \left| {\psi ^ -  } \right\rangle _{34}  \otimes \left| i \right\rangle _1  \to \left| {\psi ^ -  } \right\rangle _{41}  \otimes \left| f \right\rangle _3
\end{equation}
In this process, the partial Bell state measurement is realized by coincidence counting after a beam splitter (i.e., FBS in Fig. \ref{setup}(b)), due to the fact that only one input state, $\left| {\psi^-}\right\rangle$, out of the four Bell states has coincidence counts after the beam splitter.
The state of the photons in ch1 is heralded by the photon states in ch2  with the correlation of
$\left| {\psi ^ -  } \right\rangle _{12}  = \frac{1}{{\sqrt 2 }}(\left| {HV} \right\rangle  - \left| {VH} \right\rangle )_{12}  = \frac{1}{{\sqrt 2 }}(\left| {AD} \right\rangle  - \left| {DA} \right\rangle )_{12}$, where,
$ \left| D \right\rangle  = \frac{1}{{\sqrt 2 }}(\left| H \right\rangle  + \left| V \right\rangle ) $  (corresponding to $ \theta  = 45^ \circ$)  and
$\left| A \right\rangle  = \frac{1}{{\sqrt 2 }}(\left| H \right\rangle  - \left| V \right\rangle )$  (corresponding to $ \theta  = 135^ \circ$).

\begin{figure*}[tbp]
\includegraphics[width= 0.95 \textwidth]{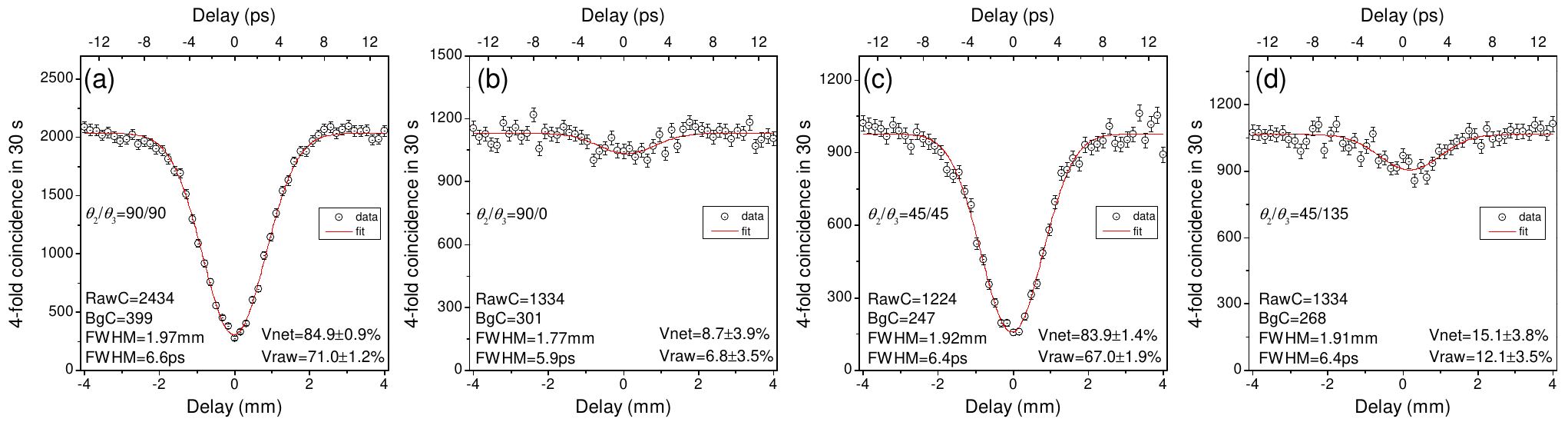}
\caption{ Teleportation with Polarizers 1/4 removed, and Polarizers 2/3 at $90^\circ$/$90^\circ$, $90^\circ$/$0^\circ$, $45^\circ$/$45^\circ$ and $45^\circ$/$135^\circ$ for (a-d), respectively. The uncertainties of these visibilities were derived using Poissonian errors on the coincidence counts. }
\label{teleportation}
\end{figure*}
\begin{table}[tbp]
\begin{ruledtabular}
\begin{tabular}{cccc}
  $\theta_2$ / $\theta_3$  & $\text{V}_{\text{net}}$ ($\text{V}_{\text{raw}}$)      & $\theta_2$ / $\theta_3$   & $\text{V}_{\text{net}}$ ($\text{V}_{\text{raw}}$)  \\
 \hline \\
  $0^\circ$/$0^\circ$              & 75.8 $\pm$ 2.6\%  (55.9 $\pm$ 3.2\%)       & $45^\circ$/$45^\circ$              & 83.9 $\pm$ 1.4\% (67.0 $\pm$ 1.9\%)   \\
  $0^\circ$/$90^\circ$              & 14.2 $\pm$ 4.1\% (11.4 $\pm$ 3.8\%)      & $45^\circ$/$135^\circ$              & 15.1 $\pm$ 3.8\% (12.1 $\pm$ 3.5\%)   \\
  $90^\circ$/$0^\circ$              & 8.7 $\pm$ 3.9\%  (6.8 $\pm$ 3.5\%)       & $135^\circ$/$45^\circ$              & 16.9 $\pm$ 3.5\% (13.2 $\pm$ 3.2\%)    \\
  $90^\circ$/$90^\circ$             & 84.9 $\pm$ 0.9\% (71.0 $\pm$ 1.2\%)       & $135^\circ$/$135^\circ$              & 81.9 $\pm$ 1.3\% (65.9 $\pm$ 1.7\%)   \\
\end{tabular}
\end{ruledtabular}
\caption{\label{teleportable} Quantum teleportation at different angles of Polarizer 2 ($\theta_2$) and Polarizer 3 ($\theta_3$). }
\end{table}

Firstly, we demonstrate a teleportation in H/V bases, as shown in Fig. \ref{teleportation}(a) and (b).
The initial state of the photons in ch1 is in H polarization, i.e., $\left| i \right\rangle _1  = \left| H \right\rangle _1$, which is heralded by their daughter photons in ch2 with V polarization ($\theta _2  = 90^ \circ$).
Then, The partial Bell state measurement on ch1 and ch4 projects the correlated photons in ch3 with H polarization, i.e., $\left| f \right\rangle _3  = \left| H \right\rangle _3$.
With this condition, if the angle of Polarizer 3 is $\theta _3  = 0^ \circ$, the 4-fold CC exists, therefore, no HOM dip  appears  at the zero delay point, as shown in Fig. \ref{teleportation}(b).
Otherwise, if  $\theta _3  = 90^ \circ$, all the H-polarized photons are blocked, then ideally, there is no 4-fold CC at the zero delay point, so a HOM dip occurs, as shown in Fig. \ref{teleportation}(a).
Similarly, we also showed the teleportation results with other bases, as shown in Fig. \ref{teleportation}(c) and (d) for  ch1 at $ \left| D \right\rangle $ bases (with $\theta _2  = 45^ \circ$).
More results are summarized in Tab. \ref{teleportable}.
The visibilities in H/H, V/V, A/A, D/D bases range from 75.8\% to 84.9\%,  all well beyond the classical limit of 50\%.

From the coincidence counts in Fig. \ref{teleportation}(a-d), we can see the SNSPDs are strongly polarization-dependent.
Because of its special construction structure,  the SNSPD has a maximal efficiency on certain polarization direction and its orthogonal direction has the minimal efficiency.
According to our experimental tests, the maximal efficiency is typically two times of the minimal efficiency.
To avoid the polarization-dependency of the SNSPDs for the entanglement swapping test, we change the FBS  in Fig. \ref{setup}(b)  to the combination of a FBS and a FPBS  in Fig. \ref{setup}(c).

\subsection{Entanglement swapping}

The principle of entanglement swapping can be understood from the following equation \cite{Pan2012}:
\begin{equation}\label{eq2}
\left| {\psi ^ -  } \right\rangle _{12}  \otimes \left| {\psi ^ -  } \right\rangle _{34}  = \frac{1}{2}(\left| {\psi ^ +  } \right\rangle _{14}  \otimes \left| {\psi ^ +  } \right\rangle _{23}  - \left| {\psi ^ -  } \right\rangle _{14}  \otimes \left| {\psi ^ -  } \right\rangle _{23}  - \left| {\phi ^ +  } \right\rangle _{14}  \otimes \left| {\phi ^ +  } \right\rangle _{23}  + \left| {\phi ^ -  } \right\rangle _{14}  \otimes \left| {\phi ^ -  } \right\rangle _{23} )
\end{equation}
The detection of an entangled  state in ch1 and ch4 heralds the existence of entanglement in ch2 and ch3, which  originally  have no correlation.
The partial Bell state measurement in  Fig. \ref{setup}(c) is realized by the combination of FBS and FPBS.
Only one input state $\left| {\psi ^ +  } \right\rangle$ out of the four Bell states has coincidence counts at port 7 and port 8 in   Fig. \ref{setup}(c), due to the transformation of a BS: $\left| {\psi ^ +  } \right\rangle _{14}  = \frac{1}{{\sqrt 2 }}(\left| {H_1 V_4 } \right\rangle  + \left| {V_1 H_4 } \right\rangle ) \to \frac{1}{{\sqrt 2 }}(\left| {H_5 V_5 } \right\rangle  - \left| {V_6 H_6 } \right\rangle )$.

To realize such a scheme in experiment, we need to firstly calibrate the photon polarizations in the FBS and FPBS.
We reinsert the two PBSs into Polarizer 1 and  Polarizer  4,  and rotate the angles of HWPs and QWPs in ch1 and ch4, so as to achieve the following condition:
the H polarized photons in ch1 travel to outport 8, while the H polarized photons in ch4 travel to outport 7.
Under this condition, the H (V) polarized photons in ch4 are converted to V (H) polarized, while the polarization of photons in ch1 is not changed and hence can function as a reference.
Therefore, an input state of $\left| {\phi ^ \pm   } \right\rangle _{14}$  is transformed to the state of $\left| {\psi ^ \pm   } \right\rangle _{14}$, and the $\left| {\psi ^ \pm   } \right\rangle _{14}$ sate is transform to $\left| {\phi ^ \pm   } \right\rangle _{14}$ state, respectively.
As a result, the state in Eq. (\ref{eq2}) is transformed to Eq. (\ref{eq3}):
\begin{equation}\label{eq3}
\left| {\psi ^ -  } \right\rangle _{12}  \otimes \left| {\psi ^ -  } \right\rangle _{34}  \to  \frac{1}{2}(\left| {\phi ^ +  } \right\rangle _{14}  \otimes \left| {\psi ^ +  } \right\rangle _{23}  - \left| {\phi ^ -  } \right\rangle _{14}  \otimes \left| {\psi ^ -  } \right\rangle _{23}  - \left| {\psi ^ +  } \right\rangle _{14}  \otimes \left| {\phi ^ +  } \right\rangle _{23}  + \left| {\psi ^ -  } \right\rangle _{14}  \otimes \left| {\phi ^ -  } \right\rangle _{23} )
\end{equation}
Let us focus on the third term. The  $\left| {\psi ^ +  } \right\rangle _{14}$  state will be detected by  the Bell state analyzer, and thus projects the state in ch2 and ch3 to $\left| {\phi ^ +  } \right\rangle _{23}$ state.

\begin{figure}[tbp]
\includegraphics[width= 0.75 \textwidth]{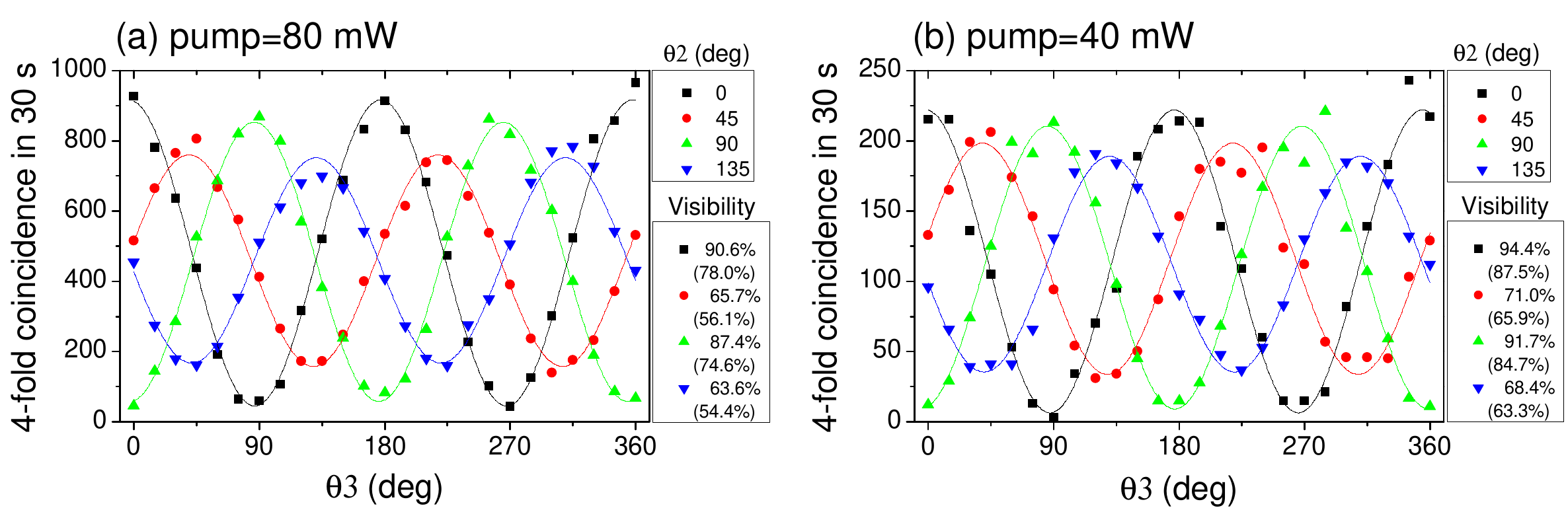}
\caption{ The entanglement swapping result. (a) with 80 (85) mW pump power for entangled source I (II). (b) the power was reduced to 40 (42.5) mW for entangled source I (II).  } \label{swap}
\end{figure}

After the calibration, the two PBSs in Polarizer 1 and  Polarizer 4 are removed.
We fix  the optical path delay at the zero delay position, then we rotate $\theta _2$/$\theta _3$ and record the 4-fold CC, whose result is shown in  Fig. \ref{swap} (a).
Figure \ref{swap} (a) shows  an experimental interference pattern of $\left| {\phi ^ +  } \right\rangle _{23}$ state, which is consistent with our theoretical expectation in Eq. (\ref{eq3}).
The net (raw) visibilities at  $\theta _2= 0^ \circ  /45^ \circ  /90^ \circ  /135^ \circ$ are 90.6\% (78.0\%)/65.7\% (56.1\%)/87.4\% (74.6\%)/63.6\% (54.4\%), respectively.
The background counts are subtracted using the same method as described in the four-fold HOM interference.

To decrease the effect of multi-photon emission,  we decrease the pump power in Fig. \ref{swap}(b) to half of the power in Fig. \ref{swap}(a), i.e., the pump power is 40 mW for Source I and 42.5 mW for Source II  in Fig. \ref{setup} (b).
We find the raw visibilities at 0$^{\circ}$/45$^{\circ}$/90$^{\circ}$/135$^{\circ}$ in Fig. \ref{swap}(a) were improved by 9.5 \%,  9.8 \%,  10.1 \% and  8.9 \% from Fig. \ref{swap}(a).

All the visibilities in Fig. \ref{swap}(a, b) are higher than 1/3, verifying the photons are entangled, according to the Peres criteria \cite{Peres1996}.
The  minimal visibility (V) in  Fig. \ref{swap}(b) is 68.4\%, corresponding to a minimal fidelity (F) of 76.3\%, by considering the relation $4F=3V+1$ \cite{Riedmatten2005}.

\section{Discussion}

\emph{Comparison of brightness}
We achieved a four-fold coincidence count rate of around 100 cps,  and  a raw (net) visibility of 73.3 $\pm$ 1.0 \% (85.1 $\pm$ 0.8\%) in the experiment.
This count rate is 3 orders higher than the previous experiments at telecom wavelength \cite{Marcikic2003, Riedmatten2005, Halder2007, Takesue2009, Aboussouan2010, Xue2010, Xue2012, Wu2013,Bruno2014} .
We compare the brightness of our result with the previous ones at telecom wavelengths in Tab. \ref{brightnesstable}.
\begin{table}[tbp]
\begin{ruledtabular}
\begin{tabular}{ccccccccc}
  Reference                                   & Group    & wavelength       & source        & 4-fold CC    & V-raw               & V-net              & qubit type  & application \\
 \hline \\
 Marcikic2003\cite{Marcikic2003}              & Geneva   &1310/1550 nm     &  LBO           & 0.05 cps     & V$_{telep.}$=70\%   & NA                  & time-bin &teleportation\\
  Riedmatten2005\cite{Riedmatten2005}         & Geneva   &1310/1550 nm     &  LBO           & 0.0037 cps   & V$_{swap.}$=80\%    & NA                    & time-bin &swapping \\
  Halder2007\cite{Halder2007}                 & Geneva   &1560 nm          &  PPLN-WG       & 0.0003 cps   & V$_{HOM}$=77\%      & NA                & time-bin &swapping\\
  Takesue2009\cite{Takesue2009}               & Atsugi   &1551 nm          &  fiber         & 0.038 cps    & V$_{HOM}$=64\%    & NA                 & time-bin &swapping\\
  Xue2012\cite{Xue2010, Xue2012}              & Tsukuba  &1538/1562 nm     &  PPLN-WG       & 0.016 cps    & V$_{HOM}$=75\% & NA                  & polarization &swapping\\
  Wu2013\cite{Wu2013}                         & Tokyo    &1550 nm          &  PPLN          & 0.08 cps     & V$_{HOM}$=NA        & 92\%                 & polarization &swapping\\
  \textbf{This work}    & \textbf{Tokyo}     &\textbf{1584 nm}   & \textbf{PPKTP}   & \textbf{108 cps}  & \textbf{V$_{HOM}$=73\%}   & \textbf{85\%}    & \textbf{polarization} & \textbf{swap.}/\textbf{telep.}\\     %
\end{tabular}
\end{ruledtabular}
\caption{\label{brightnesstable} Comparison of wavelength, photon source, four-fold  coincidence count rate (4-fold CC), raw visibility (V-raw) and net visibility (V-net) with the previous teleportation or entanglement swapping experiments at telecom wavelengths. V$_{telep.}$, V$_{swap.}$  and V$_{HOM}$  are the visibilities in teleportation, entanglement swapping and Hong-Ou-Mandel interference tests, respectively.}
\end{table}
There are mainly three reasons for this big technical jump in our high four-fold coincidence count rates.
The first one attributes to the intrinsic high spectral purity of our source.
Thanks to the group-velocity matching condition, the intrinsics purity is as high as 0.82, which is much higher than the conventional purity of PPLN crystal.
Therefore, there is no need for narrow bandpass filters, which are widely used in conventional scheme and decrease the brightness of the source severely.
The second reason comes from the optimization of the alignment, especially, the improvement of the coupling efficiency to the single mode fibers for both the clockwise pump and counter clockwise pump in the Sagnac-loop.
The last and also the most important reason is the high efficiency of our SNSPDs, which showed 30 times higher count rates than the traditional InGaAs avalanche photodiode (APD) \cite{Jin2013SNSPD}.

\emph{Application for field test in free space and fibers}
We noticed that our count rates and visibilities are comparable to the previous teleportation and entanglement swapping experiments over 100 km free space channel at $\sim$ 800 nm wavelengths \cite{Yin2012, Ma2012, Herbst2014}, as compared in  Tab. \ref{NIRbirghtness}.  Therefore, our scheme is directly applicable to the long-distance field test of teleportation and entanglement swapping at telecom wavelengths, which is the heart of a global quantum internet \cite{Kimble2008, Herbst2014, Khalique2014}.
Although conventional systems using BBO crystals and Si-APDs are applicable to free space communications, the BBO sources operating at 800 nm band are never applied  to fiber communications.
With high count rates and visibilities, our source can demonstrate the practical quantum communications using fiber infrastructures.
It is also possible to combine both the free space links and fiber links using our scheme.

\emph{Application for 6, 8-photon entangled state}
The brightness of our photon source is also comparable to the previous eight-photon entangled state generation experiments at $\sim$ 800 nm wavelengths \cite{Huang2011, Yao2012}, as  in Tab. \ref{NIRbirghtness}.  So our source can also be expanded to generate the 6, 8-photon entangled state at telecom wavelengths.
\begin{table}[tbp]
\begin{ruledtabular}
\begin{tabular}{ccccccccc}
  Reference                                   & Group    & wavelength   & source      & 2(4)-fold CC  and visibility                  & application \\
 \hline \\
 Herbst2014\cite{Herbst2014}              & Vienna      &808 nm     &  BBO            & C$_2$=130 kcps, C$_4$=100 cps, V$_{swap.}$=60\%        &swapping 143 km\\
 Yin2012\cite{Yin2012}                    & Hefei       &808 nm     &  BBO            & C$_2$=440 kcps,  V$_{ent.}$=91\%,  V$_{HOM}$=60\% $^*$    &teleportation 100 km\\
 Yao2012\cite{Yao2012}                    & Shanghai    &780 nm     &  BBO            & C$_2$=310 kcps, V$_{ent.}$=94\%,  V$_{HOM}$=76\%    &8-photon ent. state\\
 Huang2011\cite{Huang2011}                & Hefei        &780 nm     &  BBO         & C$_2$=220 kcps,  V$_{ent.}$=97\%,  V$_{HOM}$=82\%   &8-photon ent. state\\
  \textbf{This work}    & \textbf{Tokyo}     &\textbf{1584 nm}   & \textbf{PPKTP}   & \textbf{ C$_2$=150 kcps,  V$_{ent.}$=98\%,  V$_{HOM}$=78\% }           & \textbf{swap./telep.}\\
\end{tabular}
\end{ruledtabular}
\caption{\label{NIRbirghtness} Comparison with  the previous entangled source at NIR  wavelengths. V$_{ent.}$ is visibility of the entangled (ent.) state in a correlation measurement. C$_{2(4)}$ is the 2(4)-fold coincidence count rate. \\ $^*$The 4-fold CC rate is C$_4$ = 2 kcps in \cite{Yin2012}.  }  
\end{table}
All the photon sources in \cite{Yin2012, Ma2012, Herbst2014, Huang2011, Yao2012} were pumped by the second harmonic of the fundamental Ti:Sapphire lasers operating at $\sim$ 76 MHz repetition rate.
In contrast, our photon source doesn't require a second-harmonic process, which has a limited efficiency.
Therefore, our scheme is possible to achieve higher count rate, especially when pumped by high-repetition rate lasers \cite{Jin2014SPDC10GHz}.

\emph{Application for quantum key distribution (QKD)}
Our setup opens the way to practical implementation of the qubit-amplifier based  device-independent-QKD scheme, which was proposed by Gisin and colleagues in 2010 \cite{Gisin2010}.
Further, our experiment has the potential to realize the entanglement swapping based QKD (ES-QKD) protocols to double the transmission rate of QKD \cite{Waks2002, Scherer2011}.
With 4-fold CC rate of over 100 cps, one can accumulate sufficient size of sift key for a reasonable span of time even in practical lossy environment.
For example, with a total system loss of 10 dB (50 km distance in standard fiber), the 4-fold CC rate can still remain 10 cps.
One can then collect large enough data within a system stability time scale, and validate the system performance, namely the precise evaluation of the Bell inequality and the secure key rate.

\emph{Application for quantum repeater}
In the scenario of quantum repeaters, a point to point quantum communication between remote locations is limited to about 300-500 km due to the losses in fibers, but this  problem can be solved by decompose the long distance into serval shorter elementary links.
In each link, the entanglement is shared and stored in quantum memories with long coherence time.
Finally, the entangled state is retrieved from the quantum memories on demand and swapped between adjacent nodes, so as to faithfully increase the communication distance.
Entanglement swapping is the key building block for the construction of  quantum repeaters.
The recent experimental breakthrough of quantum memory at telecom wavelength has also been reported \cite{Saglamyurek2014}.
The highly efficient entanglement swapping in this experiment will be an important experimental step toward the realization of quantum repeater protocols.

\emph{How to reduce the multi-pair emission}
Multi-photon emission is the main reason for the degradation of raw visibilities in our teleportation and entanglement swapping experiments.
To obtain high count rates, we need to excite the SPDC with high pump powers, which inevitably lead to stronger multi-photon emission in SPDC.
To solve this problem, recently, we propose and demonstrated a new method - increasing the repetition rate of the pump laser using a 10 GHz repetition rate comb laser \cite{Jin2014SPDC10GHz}.
With such a high repetition rate, we can maintain the high visibilities at high pump powers.
In this work, the limitation of the brightness of our photon source mainly comes from the low repetition rate  of the pump laser.

\section{Methods}

\noindent
\textbf{Entangled photon source with GVM condition}
Our pulsed polarization-entangled photon source is generated from a periodically poled KTiOPO$_4$ (PPKTP) crystal in a
Sagnac interferometer configuration. Since the group-velocity-matching (GVM) condition is satisfied \cite{Grice1997, Konig2004}, the intrinsic spectral
purity of the photons is much higher than the conventional schemes. Therefore, there is no need to use narrow bandpass filters to improve the spectral purity \cite{Eckstein2011, Gerrits2011}. The combination of a Sagnac interferometer and the
GVM-PPKTP crystal makes our entangled source compact, stable, highly entangled, spectrally pure and ultra-bright \cite{Jin2014OE}.
The mean photon numbers per pulse are $\sim$ 0.1 in our photon source with a pump power of $\sim$ 80 mW. The overall detecting efficiency is  $\sim$ 0.2.

\noindent
\textbf{The SNSPDs}
Our superconducting nanowire single photon detectors (SNSPDs) have a system detection efficiency (SDE) of around 70\% with a dark count rate (DCR) less than 1 kcps \cite{Miki2013, Yamashita2013, Jin2013SNSPD}.
The SNSPD also has a wide spectral response range that covers at least from 1470 nm to 1630 nm wavelengths \cite{Jin2013SNSPD}.
The measured timing jitter and  dead time (recovery time) were  68 ps \cite{Miki2013} and 40 ns \cite{Miki2007}.

\section*{Acknowledgements}
The authors thank K. Wakui, M. Fujiwara, T. Yamashita, S. Miki, H. Terai and Z. Wang for helpful discussion and assistance in experiment.
This work was supported by the Founding Program for World-Leading Innovative R\&D on Science and Technology (FIRST).

\end{document}